\documentstyle[twocolumn,prl,aps,epsf]{revtex}

\def\Vec#1{\mbox{\boldmath $#1$}}

\begin{document}
\draft
\wideabs{
\title{Rotating superfluid turbulence}
\author{Makoto Tsubota$^1$, Tsunehiko Araki$^1$ and Carlo F. Barenghi$^2$}
\address{$^1$ Department of Physics, Osaka City University,\\
Sumiyoshi-Ku, Osaka 558-8585, Japan\\
$^2$School of Mathematics and Statistics, University of Newcastle,\\
Newcastle upon Tyne NE1 7RU, UK}
\date{\today}
\maketitle
\begin{abstract}
Almost all studies of vortex states in helium~II have been concerned
with either ordered vortex arrays
or disordered vortex tangles.
This work studies numerically
what happens in the presence of both rotation (which induces order)
and thermal counterflow (which induces disorder).
We find a new statistically steady state in which the
vortex tangle is polarized along the rotational axis.
Our results are used to interpret
an instability which was discovered experimentally by Swanson {\it et al.}
years ago and the vortex state beyond the instability that has been
unexplained
until now.
\end{abstract}
}

Most configurations of quantized vortices \cite{Donnelly} which have been
investigated in helium~II can be grouped into two types: ordered vortex
arrays and disordered vortex tangles.
An ordered array is made when helium~II is rotated with
angular velocity $\Omega$ exceeding a certain small critical value.
The resulting quantized vortices are aligned along the rotation axis and
form an array whose areal number density is given by Feynman's rule
$L_{\rm  rot} =2\Omega/\kappa$,
where $\kappa=9.97 \times 10^{-4}{\rm cm^2/sec}$ is the quantum of
circulation.
A spatially disordered tangle is obtained
\cite{Vinen,Tough}
when helium is made turbulent under thermal counterflow velocity
$V_{\rm  ns}$ faster than some small critical value, or, more in
general \cite{Niemela},  using grids
\cite{Stalp}
or propellers \cite{Tabeling}. In this work we concentrate on
counterflow because this form of turbulence
is simpler (no need to worry about large scale motion
and effects induced by normal fluid eddies).
In turbulent counterflow the vortex line density
(length of vortex line per unit volume) is
$L_{\rm  flow} =\gamma_H^2 V_{\rm  ns}^2$ where $\gamma_H$
is a temperature dependent coefficient \cite{Tough}.
This vortex system is almost isotropic, provided that one neglects a small
anisotropy
induced by the imposed counterflow\cite{Wang}.

An important question naturally arises: what happens if vortices are created
by {\it both} rotation and counterflow along the rotational axis?
We are aware of only one experiment which addressed this issue, which was
performed by Swanson, Barenghi and Donnelly
\cite{Swanson} years ago.
The counterflow channel was mounted on a rotating cryostat, so it was
possible to create vortices by independent combination of rotation
and counterflow.
The absolute value of the vortex line density $L$ was determined
from the measurement of the
attenuation of a second sound along the channel, which was
calibrated against the known density in rotation \cite{BPD}.
Swanson {\it et al.} found that at slow rotation the
critical counterflow velocity
above which the flow became turbulent was greatly reduced.
Furthermore two critical velocities $V_{c1}$ and $V_{c2}$ were
observed.  For $V_{\rm  ns} < V_{c1}$
the measured vortex line density  $L=L_{\rm rot}$ was independent of
the small values of $V_{ns}$ and in agreement with Feynman's rule,
which was evidence for an ordered vortex array.
The value of $V_{c1}$ was consistent with the
critical velocity of a vortex wave instability which
had been first observed by Cheng {\it et al.} \cite{Cheng} and then
explained by Glaberson {\it et al.} \cite{Glaberson}.
This instability, hereafter referred to as the Donnelly-Glaberson(DG)
instability,
takes the form of Kelvin waves (helical displacements of the vortex cores)
which are destabilized by the component of the counterflow velocity
in the direction along the vortices.

Unfortunately, it was not clear to Swanson {\it et al.} \cite{Swanson}
what was the nature of the flow beyond the DG instability
($V_{\rm ns}>V_{c1}$).
Their experiment showed that rotation added
fewer than the expected $2\Omega/\kappa$ vortex lines to those
made by the counterflow. Lacking any direct flow visualization
the nature of the flow was a mystery. Theoretically, the
linear stability analysis of Glaberson determines only the value
$V_{c1}$ of the instability of the vortex array. What happens
beyond the instability can be studied only by nonlinear analysis.
Thus the pioneering experiment of Swanson {\it et al.}
has lacked a theoretical interpretation.

Motivated by their work, we study numerically quantized vortices
under both rotation and counterflow using the vortex filament model.
First we prepare an initial vortex array in rotation,  then we
apply a counterflow $V_{\rm ns}$ along the rotational axis.
We find that when $V_{\rm ns}>V_{c1}$
Kelvin waves are excited and grow (DG instability),
and the vortex array becomes unstable.
Then we show the first numerical evidence for a polarized vortex tangle
(polarized superfluid turbulence).

Our work is
relevant to other contexts of current interest, such as
superfluid $^3$He and atomic Bose- Einstein condensates \cite{BDV2,BEC}.
In general, our problem is concerned with the competition
between order (here induced by rotation) and disorder (here induced
by the heat flow).

For superfluid $^4$He, the vortex filament model is very useful,
because the vortex core radius $a_0 \sim 10^{-8}$ cm is microscopic
and the circulation $\kappa = 9.97 \times 10^{-4}$ cm$^2$/sec is fixed.
Neglecting the normal fluid, according to Helmholtz's
theorem, a superfluid vortex at a point moves
with velocity which depends on the shape of the vortex and on
the velocity field
$\dot{\Vec{s}}_0$ at that point which is induced by other
vortices \cite{Schwarz}.
Therefore, in order to study a rotating tangle, we need to formulate the
laws of
vortex dynamics in a rotating frame \cite{Yoneda}.

The natural way to perform the calculation in a rotating frame would
require to consider a cylindrical container. We do not
follow this approach for two reasons.
Firstly, our formulation is implemented using the
full Biot - Savart law, not the localized-induction approximation
often used in the literature.
This would require to place image
vortices beyond the solid boundary to impose the condition of no flow
across it.  This has been done in cartesian (cubic) geometry,
but it is difficult to do in  cylindrical geometry,
Secondly, the original experiment by Swanson {\it et al.} \cite{Swanson} was
carried out in a rotating channel with a square cross section.

In a rotating vessel the equation of motion of vortices is modified by two
effects.  The first effect is the force acting upon the vortex due to
the rotation.  According to  Helmholtz's theorem, the generalized
force acting upon the vortex is balanced by the Magnus force as
\begin{equation}
\rho_{\rm s}\kappa (\Vec{s}' \times \dot{\Vec{s}}_0)=\frac{\delta F'}{\delta
\Vec{s}}, \label{rot3}
\end{equation}
where $F'=F-\Vec{\Omega}\cdot \Vec{M}$ is the free energy of a system in a
frame rotating
around a fixed axis with angular velocity $\Vec{\Omega}$ and angular
momentum $\Vec{M}$.
Taking the vector product with $\Vec{s}'$, we obtain the velocity
$\dot{\Vec{s}}_0$.
The first term $F$ due to the kinetic energy of vortices gives that
Biot - Savart law, and the second
term $\Vec{\Omega}\cdot \Vec{M}$ leads to the velocity $\dot{\Vec{s}}_{\rm
rot}$
of the vortex caused by the rotation.
The second effect is the superflow induced by the rotating vessel.
For a perfect fluid we know the analytical solution of the velocity
$\Vec{v}_{\rm cub}$
inside a cube of size $D$ rotating with angular velocity
$\Vec{\Omega}=\Omega \hat{\Vec{z}}$ \cite{Thomson}.
By taking into account  both effects, we eventually obtain the velocity
$\dot{\Vec{s}}_0$ in a rotating frame which is
\begin{eqnarray}
\dot{\Vec{s}}_0 & = & \frac{\kappa}{4\pi}\Vec{s}' \times \Vec{s}'' \ln \left(
 \frac{2 (l_{+} l_{-})^{1/2}}
{e^{1/4} a_0} \right) \nonumber \\
 & {} &  + \frac{\kappa}{4\pi} \int ' \frac{(\Vec{s}_1 -
\Vec{r}) \times d\Vec{s}_1}
{|\Vec{s}_1 - \Vec{r}|^3}  + \dot{\Vec{s}}_{\rm rot} + \Vec{v}_{\rm cub} .
\label{rot7}
\end{eqnarray}
where the first term at the right hand side is the usual de-singularization
of the Biot - Savart integral which is well known in the literature.
Here the vortex filament is represented by the parametric form
$\Vec{s}=\Vec{s}(\xi,t)$.
The symbols $l_{+}$ and $l_{-}$ are the lengths of the two adjacent line
elements after discretization,
and the prime denotes differentiation with respect to the arc length $\xi$.
The second term represents the nonlocal field by carrying out the integral
along the rest of the filament on which $\Vec{s}_1$ refers to a point.

At a finite temperature, a quantized vortex is also affected by the mutual
friction, as the vortex core is dragged by the normal flow. Taking this
effect into account
the velocity of a vortex line at the point $\Vec{s}$ is given by
\begin{equation}
\dot{\Vec{s}}=\dot{\Vec{s}} _0+ \alpha \Vec{s}' \times (\Vec{v}_{\rm ns} -
\dot{\Vec{s}}_0)- \alpha '
\Vec{s}' \times [\Vec{s}' \times (\Vec{v}_{\rm ns} - \dot{\Vec{s}}_0)],
\label{rot2}
\end{equation}
where $\alpha$ and $\alpha'$ are the temperature-dependent friction
coefficients.

Two quantities are useful for characterizing the rotating tangle.
The first is the vortex line density $L=(1/\Lambda) \int d\xi$,
where the integral is taken along all vortices in the sample volume $\Lambda$.
The second is the polarization of the tangle, which we define
by $<s'_z>=(1/\Lambda L) \int d\xi \Vec{s}'(\xi)\cdot \hat{\Vec{z}}$
where $\hat{\Vec{z}}$ is the unit vector along the $z$
direction. Note that $<s'_z>$ is unity for a vortex array and zero for
a randomly oriented tangle.

We now describe briefly how to perform the numerical calculation \cite{TAN}.
A vortex filament is represented by a single string of points at a distance
$\Delta \xi$ apart.
When two vortices approach within $\Delta \xi$, they are assumed to reconnect.
The computational sample is taken to be a cube of size $D=1.0$ cm.
We adopt periodic boundary conditions along the rotating axis and rigid
boundary conditions at the side walls.
All calculations are made under the full Biot - Savart law, placing image
vortices beyond the solid boundaries.
Our space resolution is $\Delta \xi =1.83 \times 10^{-2}$ cm and the time
resolution is
$\Delta t=4.0 \times 10^{-3}$ sec.
The counterflow $\Vec{V}_{\rm ns}$ is applied along the $z$ axis.
The normal fluid is assumed to be at rest in the rotating frame, and,
to make comparison with the experiment \cite{Swanson}, we
use $\alpha=0.1$ and $\alpha'=0$ for the temperature $T=1.6$ K.

Swanson {\it et al.} \cite{Swanson} found that the first critical velocity
$V_{c1}$ is proportional to $\Omega^{1/2}$, in agreement with the
DG instability of Kelvin waves.
Glaberson {\it et al.} \cite{Glaberson} modelled a vortex array
rotating at angular velocity $\Omega$
as a continuum and found that,
in the absence of the friction, the angular velocity of the Kelvin wave
of wavenumber $k$ is given by the dispersion relation
$\omega=2\Omega +\nu k^2$,
where $\nu=(\kappa/4 \pi) \ln{(b/a_0)}$
and  $b\approx L^{-1/2}$ is
the average distance between vortices.
This dispersion law has a critical velocity
$V_{DG}=(\omega/k)_{min}=2(2\Omega \nu)^{1/2} $ at the critical wavenumber
$k_{DG}=\sqrt{2\Omega/\nu}$. Therefore,
if the axial flow $V_{\rm ns}$ exceeds $V_{\rm DG}$ for some value of $k$,
Kelvin waves with that wavenumber $k$ become unstable and
grow exponentially in time. Physically, the phase velocity of the
mode $k$ is equal to the axial flow, so energy is fed into the Kelvin wave
by the flow.

Figure 1 illustrates the DG instability.
The computation was performed with angular velocity
$\Omega=9.97 \times 10^{-3} {\rm rad/sec}$, for which
$V_{\rm DG}=0.010 {\rm cm/sec}$.
The vortex lines remain stable when $V_{ns}=0.008{\rm cm/sec}<V_{\rm DG}$
(Fig.1a),
while at $V_{ns}=0.015{\rm cm/sec}>V_{\rm DG }$ Kelvin waves become unstable
and grow (Fig.1b), as predicted.
Figure 1c shows that Kelvin waves of larger wavenumber become unstable
at higher counterflow velocity.
This is the first numerical confirmation of the DG instability in a
rotating vortex system.

\begin{figure}[tbhp]
\begin{minipage}{1.0\linewidth}
\begin{center}
\epsfxsize=1.0\linewidth \epsfbox{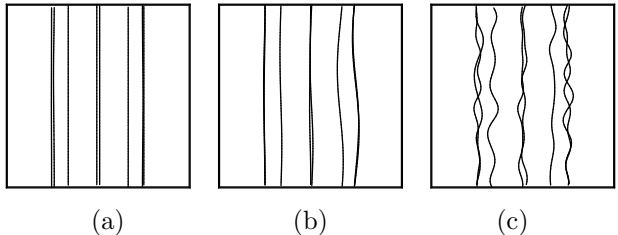}
 (a) \hspace{2cm} (b) \hspace{2cm} (c)
\end{center}
\end{minipage}
\caption{Numerical simulations of the Donnelly-Glaberson
instability at $\Omega=9.97 \times 10^{-3} {\rm rad/sec}$,
$T=1.6{\rm K}$.
Each snapshot of the vortex configuration is for the counterflow velocity
$V_{ns}=0.008{\rm cm/sec}$ (a), $0.015{\rm cm/sec}$ (b) and
$0.05{\rm cm/sec}$ (c).}
 \label{rot1}
\end{figure}

The key question is: what happens to the vortices beyond the DG instability?
Because of the computational cost of the Biot - Savart law, it is
not practically possible to compute vortex tangles with densities
which are as high ($L={\cal O}(10^4){\rm cm^{-2}}$)
as those achieved in the experiment.
Nevertheless the following numerical simulations can shed light into
the physical processes involved.
The time sequence contained in  Fig. 2 illustrates the evolution
of a  vortex array at $\Omega=4.98 \times 10^{-2}{\rm rad/sec}$
under the counterflow $V_{ns}=0.08{\rm cm/sec}$.
Figure 2a shows the initial $N=33$
parallel vortex lines; they have been seeded with
small random perturbations to make the simulation more realistic.
As the evolution proceeds, perturbations with  high wavenumbers
are damped by friction, whereas perturbations which are linearly DG-unstable
grow exponentially, hence Kelvin waves become visible (Fig. 2b).
When the amplitude of the Kelvin waves becomes of the order of the average
vortex separation, reconnections take place (Fig. 2c).
The resulting vortex loops disturb the initial vortex array, leading to an
apparently random vortex tangle (Fig. 2d).
After the initial exponential growth (which is predicted by the theory of the
DG instability), nonlinear effects (vortex interactions and vortex
reconnections) become important and nonlinear saturation takes
place. The two effects are apparent in Figure 3a, which shows the initial
exponential growth of $L$ (linear instability) and the successive
equilibration to a statistical steady state (nonlinear saturation).

Looking carefully at the saturated tangle in Fig. 2d
we notice that there are more loops oriented vertically than horizontally.
Figure 3b shows how the polarization $<s'_z>$ changes with time in the
calculation
presented in Fig.2. During the exponential growth (linear phase)
$<s'_z>$ decreases from unity (ordered vortex array), but it never
becomes zero (random tangle),
settling instead to the finite value $<s'_z>\approx 0.5$ as soon as nonlinear
saturation takes place at about $t \approx 100\rm sec$.

\begin{figure}[tbhp]
\begin{minipage}{1.0\linewidth}
\begin{center}
\epsfxsize=1.0\linewidth \epsfbox{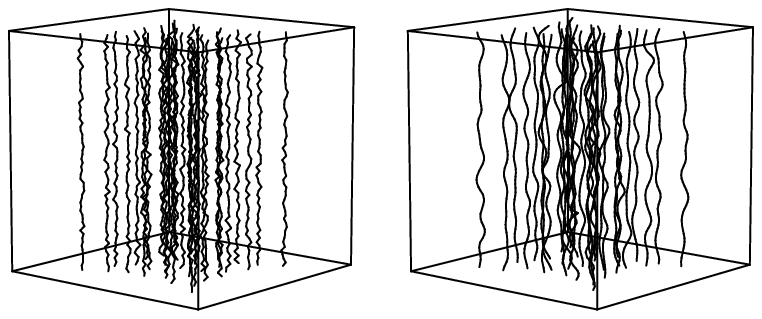}
(a) \hspace{3cm} (b)
\end{center}
\end{minipage}

\begin{minipage}{1.0\linewidth}
\begin{center}
\epsfxsize=1.0\linewidth \epsfbox{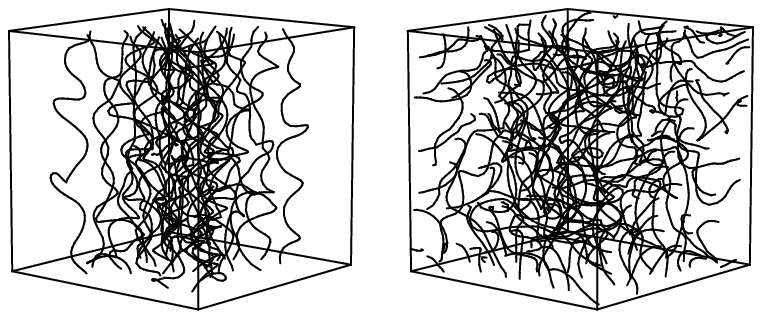}
(c) \hspace{3cm} (d)
\end{center}
\end{minipage}
\caption{Numerical simulation of rotating turbulence at
$T=1.6{\rm K}$, $\Omega=4.98 \times 10^{-2}{\rm rad/sec}$
and $V_{ns}=0.08{\rm cm/sec}$. Computed vortex tangle at
the following times:
(a): t=0{\rm sec};
(b): t=12{\rm sec};
(c): t=28{\rm sec};
(d): t=160{\rm sec}.}
 \label{rot2}
\end{figure}

\begin{figure}[tbhp]
\begin{minipage}{1.0\linewidth}
\begin{center}
\epsfxsize=1.0\linewidth \epsfbox{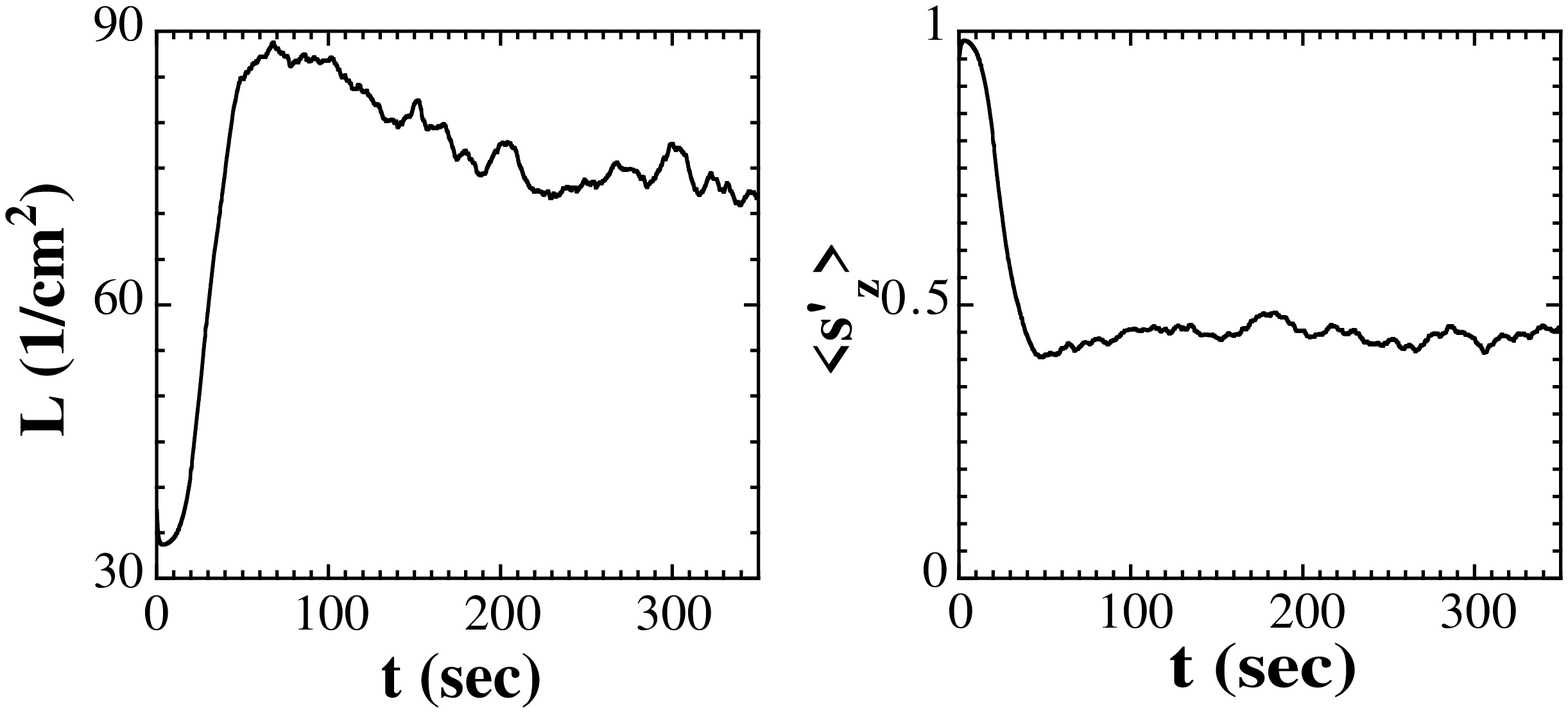}
(a) \hspace{3cm} (b)
\end{center}
\end{minipage}
\caption{Time evolution of $L$ (a) and $<s'_z>$ (b) at $T=1.6{\rm K}$ and
$V_{ns}=0.08{\rm cm/sec}$
 for $\Omega=4.98\times 10^{-2}{\rm rad/sec}$.}
 \label{rot3}
\end{figure}

Figure 4 shows the calculated dependence of the vortex line density $L$
on the counterflow velocity $V_{ns}$ at different rotation rates $\Omega$.
The dependence of $L$ on $V_{ns}$ is similar to
what appears
in the Figure 1 of the paper by Swanson {\it et al} \cite{Swanson}.
The only difference is that the scale of the axes in the paper by
Swanson {\it et al.} is bigger - in this particular figure they report
vortex line densities as high as $L \approx 2500\rm cm^{-2}$, whereas
our calculations are limited to $L\approx 80\rm cm^{-2}$.
Despite the lack of overlap between the experimental and numerical ranges,
there is clear qualitative similarity between the figures.
It is apparent that the critical velocity beyond which $L$ increases
with $V_{ns}$ is much reduced by the presence of rotation,
which is consistent with the experiment.
Figure 5 shows the calculated polarization $<s'_z>$ as a function of
counterflow velocity $V_{ns}$ at different rotation rates $\Omega$.
It is apparent that the polarization decreases with the counterflow
velocity and increases with the rotation.

\begin{figure}[tbhp]
\begin{center}
\epsfxsize=0.8\linewidth \epsfbox{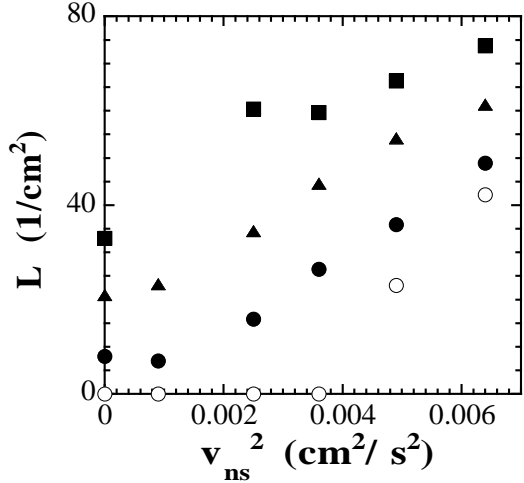}
\end{center}
\caption{Vortex line density $L$ vs $V_{ns}^2$
at $T=1.6{\rm K}$ for $\Omega=0$ (write circle), $\Omega=9.97 \times
10^{-3}{\rm rad/sec}$ (black circle),
  $\Omega=2.99 \times 10^{-2}{\rm rad/sec}$ (triangle) and
$\Omega=4.98\times 10^{-2}{\rm rad/sec}$ (square).}
 \label{rot4}
\end{figure}

In conclusion, we have studied numerically vortex tangles under the
effect of rotation for the first time. At velocities higher than the
onset of the DG instability we have determined the existence
of a new state of superfluid turbulence (polarized turbulence)
which is characterized by two statistically steady state properties,
the vortex line density and the degree of polarization. Although
the computed range of vortex line densities does not overlap with the
much higher values obtained in the experiments, we find the same
qualitative dependence of vortex line density versus counterflow
velocity at different rotations.

Further work will investigate other aspects of the problem,
particularly the nature of $V_{c2}$ and what happens
at very high counterflow velocities.
We also hope that this work will stimulate more
experiments. For example, it should be possible to observe the
polarization of turbulence by using simultaneous measurements of
second sound attenuation along and across the rotation axis.


The authors thank W.F. Vinen for useful discussions.
C.F.B. is grateful to the Royal Society for financially supporting this
project.
M.T. is grateful to Japan Society for the Promotion of Science for
financially supporting this Japan-UK Scientific Cooperative Program(Joint
Research Project).

\begin{figure}[tbhp]
\begin{center}
\epsfxsize=0.8\linewidth \epsfbox{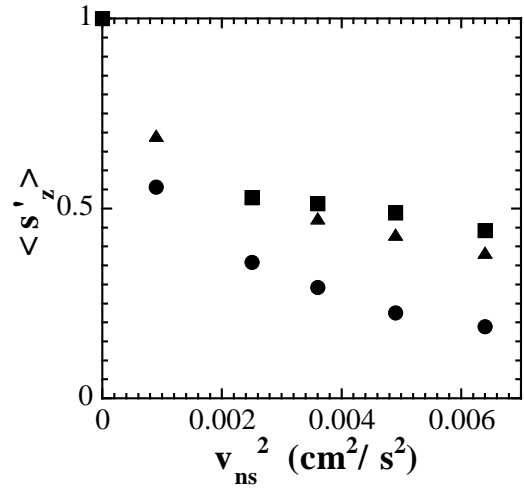}
\end{center}
\caption{Tangle's polarization $<s'_z>$ vs $V_{ns}^2$
 at $T=1.6{\rm K}$ for $\Omega=9.97 \times 10^{-3}{\rm rad/sec}$ (circle),
  $\Omega=2.99 \times 10^{-2}{\rm rad/sec}$ (triangle) and
$\Omega=4.98\times 10^{-2}{\rm rad/sec}$ (square).}
 \label{rot5}
\end{figure}

\end{document}